\documentclass[10pt,letterpaper,british,10pt]{article}
\usepackage[T1]{fontenc}
\usepackage[latin9]{inputenc}
\usepackage{amsmath}
\usepackage{amssymb}
\usepackage{graphicx}

\makeatletter

\usepackage{spconf}

\usepackage{url}

\makeatother

\usepackage{babel}
\begin{document}
\ninept\name{Jean-Marc Valin
\thanks{\copyright 2008 IEEE.  Personal use of this material is permitted. Permission from IEEE must be obtained for all other uses, in any current or future media, including reprinting/republishing this material for advertising or promotional purposes, creating new collective works, for resale or redistribution to servers or lists, or reuse of any copyrighted component of this work in other works.
}}
\address{CSIRO ICT Centre, Sydney, Australia (jmvalin@jmvalin.ca)}

\title{Perceptually-Motivated Nonlinear Channel Decorrelation For Stereo
Acoustic Echo Cancellation}
\maketitle
\begin{abstract}
Acoustic echo cancellation with stereo signals is generally an under-determined
problem because of the high coherence between the left and right channels.
In this paper, we present a novel method of significantly reducing
inter-channel coherence without affecting the audio quality. Our work
takes into account psychoacoustic masking and binaural auditory cues.
The proposed non-linear processing combines a shaped comb-allpass
(SCAL) filter with the injection of psychoacoustically masked noise.
We show that the proposed method performs significantly better than
other known methods for reducing inter-channel coherence.
\end{abstract}

\section{Introduction}

As videoconferencing applications incorporate higher sampling rates
and multiple channels, the problem of cancelling acoustic echo becomes
harder. One of the main difficulties in stereo echo cancellation is
the strong coherence that exists between the left and right channel,
making it hard or even impossible to correctly estimate the acoustic
impulse response \cite{Benesty1998}. 

\begin{figure}
\begin{center}\includegraphics[width=1\columnwidth,keepaspectratio]{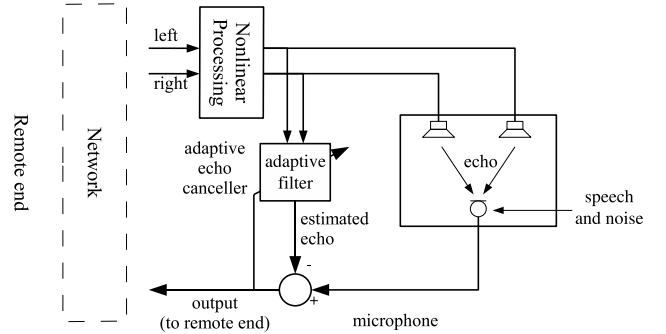}\end{center}

\vspace{-4mm}

\caption{Stereo echo cancellation system.\label{fig:Stereo-echo-cancellation} }

\vspace{-3mm}
\end{figure}

To improve the performance of a stereo acoustic echo canceller, it
is very useful to reduce the coherence between channels \cite{Benesty1998,Sondhi1995}.
This can be done by altering the signals using some form of non-linear
transformation, as illustrated in Fig. \ref{fig:Stereo-echo-cancellation}.
However, most of the methods proposed so far to reduce inter-channel
coherence \cite{Morgan2001,Ali1998,WU2005} do not take into account
human perception and tend to introduce too much audible distortion
to the signal, especially at high sampling rates. In this paper, we
propose a non-linear processing that closely matches human perception
in order to minimise coherence while maximising audio quality. We
show that by combining a shaped comb-allpass filter and psychoacoustically
masked noise, it is possible to achieve better reduction in coherence
while preserving higher audio quality than other nonlinear algorithms.

The paper is divided as follows. Section \ref{sec:Overview-And-Motivations}
presents an overview of the stereo acoustic echo cancellation problem.
Sections \ref{sec:Shaped-Comb-Allpass-Filtering} and \ref{sec:Psychoacoustically-Masked-Noise}
describe the two parts of the algorithm, which are the all-pass filtering
and the noise injection, respectively. Section \ref{sec:Evaluation-And-Results}
presents comparative results and Section \ref{sec:Conclusion} concludes
this paper.

\section{Overview}

\label{sec:Overview-And-Motivations}

In a multi-channel audio system, there usually exists a strong coherence
between channels (loudspeakers) that causes the filter optimisation
problem to be ill-conditioned. It is shown in \cite{Gansler2002}
that the normalised misalignment $\eta(n)$ is inversely proportional
to $\left(1-\gamma^{2}\right)$, where $\gamma$ is the inter-channel
coherence. In the frequency domain, the square inter-channel coherence
is defined as \cite{Benesty1998}
\begin{equation}
\gamma^{2}\left(f\right)=\frac{\left|E\left\{ X_{1}^{*}\left(f\right)X_{2}\left(f\right)\right\} \right|^{2}}{E\left\{ \left|X_{1}\left(f\right)\right|^{2}\right\} E\left\{ \left|X_{2}\left(f\right)\right|^{2}\right\} }\ ,\label{eq:coherence}
\end{equation}
where $X_{j}\left(f\right)$ denotes the Fourier transform of channel
$j$, $E\left\{ \cdot\right\} $ denotes the mathematical expectation,
and $\left(\cdot\right)^{*}$ denotes the complex conjugate.

It is desirable to maximise the audio quality, while minimising the
inter-channel coherence. Several approaches proposed so far to reduce
inter-channel coherence have focused on using memoryless non-linearities
\cite{Morgan2001}. Although they have the main advantage of being
easy to compute, these non-linearities introduce a great amount of
inter-modulation distortion, which quickly degrades the audio quality.
They also provide little control regarding how much perturbation is
caused as a function of frequency.

Another popular approach is to alter the phase of the signal in a
time varying way \cite{Ali1998,WU2005}. The time-varying aspect of
the transformation is important because the transformation would otherwise
be linear and thus have no effect on inter-channel coherence. The
phase of an audio can be altered either through the use of an all-pass
filter, or in the short-term Fourier transform (STFT) domain.

The algorithm we propose in this work was designed to minimise inter-channel
coherence, while maintaining good quality audio, including the stereo
image. Additionally, it is important not to add any significant delay
to the audio because latency is a very important aspect in the perception
of acoustic echo. Although this rules out analysis/synthesis algorithms
based on the DFT, it still allows the use overlapping windows, as
long as the processing within each window is causal. 

\begin{figure}
\begin{center}\includegraphics[width=0.65\columnwidth,keepaspectratio]{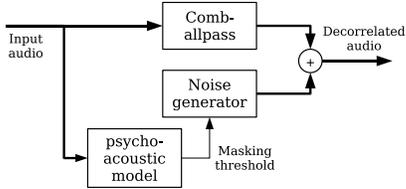}\end{center}

\vspace{-4mm}

\caption{Overview of the proposed algorithm (for each channel).\label{fig:Overview-of-system}}

\vspace{-3mm}
\end{figure}

The human auditory system localises sounds using two sets of binaural
cues. Interaural phase difference (IPD) is used at low frequencies
($\leq1.5\:\mathrm{kHz}$) and while interaural intensity difference
(IID) is used at higher frequencies ($\geq2\:\mathrm{kHz}$). We propose
a strategy that takes into account these binaural cues when altering
the audio signals. At higher frequencies, we propose to reduce coherence
by altering the phase (IPD) while preserving the IID. Because it is
not possible to alter IID without altering IPD at lower frequencies,
we use wideband psychoacoustically-masked noise with emphasis on lower
frequencies. Note that, unlike \cite{Benesty1998}, the noise is not
only shaped, but predominantly added to the lower frequencies. The
approach is illustrated in Fig. \ref{fig:Overview-of-system}.

\section{Shaped Comb-Allpass (SCAL) Filtering}

\label{sec:Shaped-Comb-Allpass-Filtering}

Allpass filters have a flat frequency response with non-linear phase
and can be represented by the general causal form
\begin{equation}
A(z)=\frac{\sum_{k=1}^{N}a_{k}z^{k-N}+z^{-N}}{1-\sum_{k=1}^{N}a_{k}z^{-k}}\ .\label{eq:allpass-causal}
\end{equation}
It is hard to design an all-pass filter starting from the general
form (\ref{eq:allpass-causal}) for high orders. However, it is possible
to construct a filter that alters the phase similarly across all frequencies
by using a simple comb-allpass filter of the form
\begin{equation}
A(z)=\frac{\alpha+z^{-N}}{1-\alpha z^{-N}}\ .\label{eq:comb-allpass}
\end{equation}
The filter in (\ref{eq:comb-allpass}) combines an all-pole comb filter
to a maximum-phase all-zero comb filter, so the poles and zeros are
equally spread around the unit circle with radii of respectively $\alpha^{1/N}$
and $\alpha^{-1/N}$.

For the processing to be non-linear, it is required to vary the coefficient
$\alpha$ controlling the filter. This is achieved through using overlapping
windows with $\alpha$ held constant over each window. We use both
an analysis window and a synthesis window to prevent any blocking
artifacts. The signal is reconstructed using weighted overlap-add
(WOLA). Because all-pass filtering is a time-domain process, no extra
delay is added because at any given time, we do not need to apply
the allpass filter on the whole window. The analysis-synthesis window
is required to meet the Princen-Bradley criterion \cite{Princen1986}
and we use the Vorbis window \cite{VorbisSpec}.

\begin{figure}
\begin{center}\includegraphics[width=0.44\columnwidth,keepaspectratio]{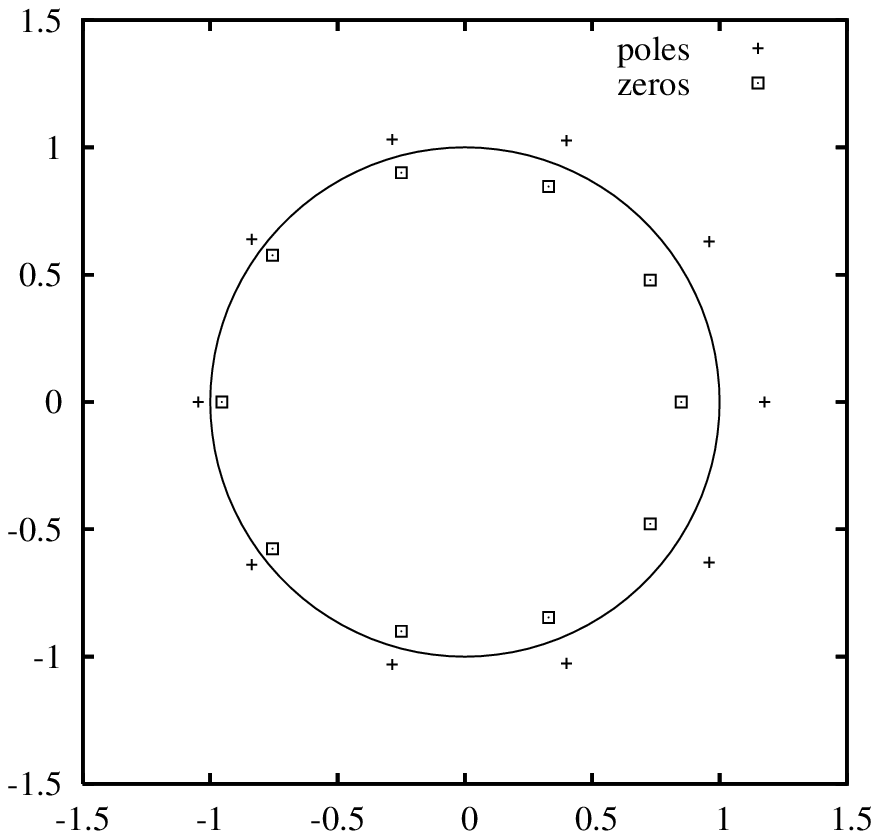}\includegraphics[width=0.56\columnwidth,keepaspectratio]{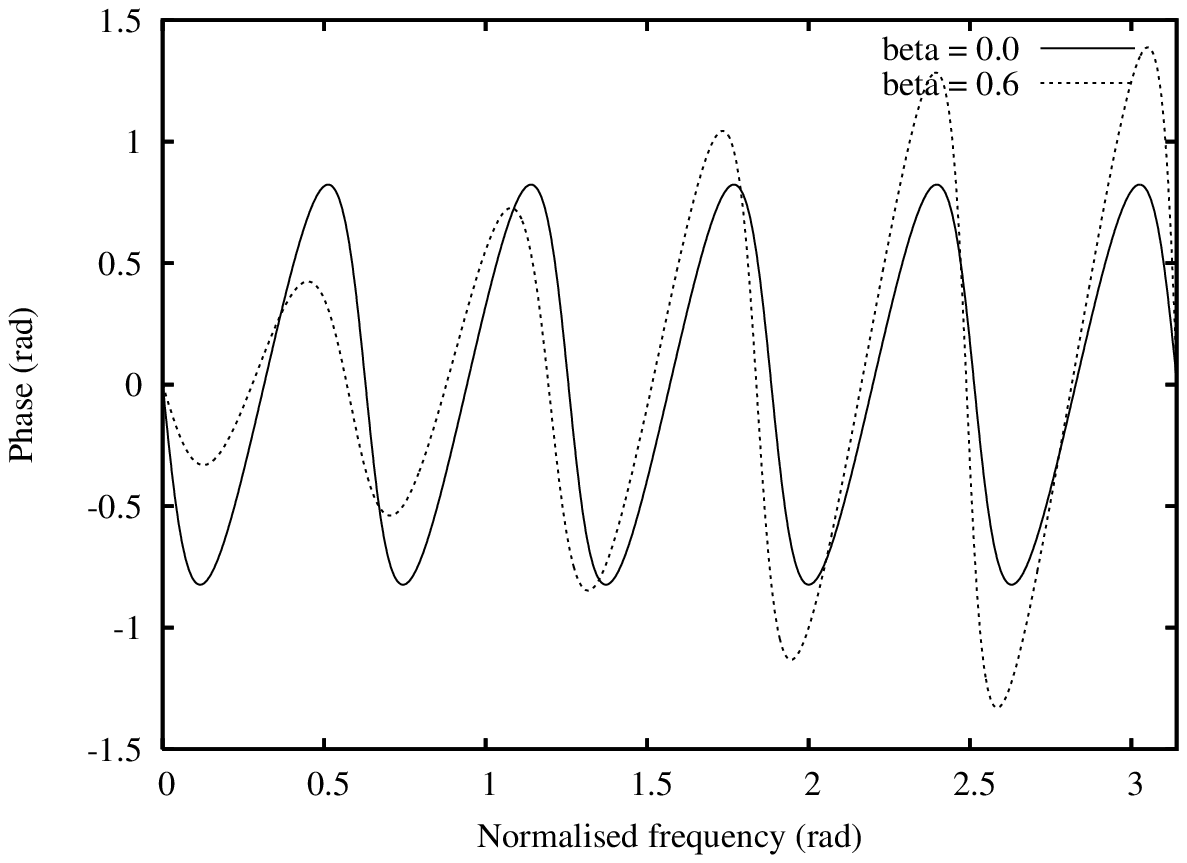}\end{center}

\vspace{-4mm}

\caption{Effect of the \emph{tilt} parameter $\beta$ (with $\alpha=0.4$,
$N=10$). Location of the poles and zeros (left) and phase response
for different values of $\beta$ (right).\label{fig:Effect-of-tilt} }
\vspace{-2mm}
\end{figure}

Interaural phase difference (IPD) is the main localisation cue at
lower frequencies, so the human ear is more sensitive to phase distortion
in the low frequencies. For that reason, it is important to ``shape''
the phase modulation as a function of frequency so as to limit distortion
of the stereo image. It is desirable to introduce less distortion
to the phase at lower frequencies than at higher frequencies. To do
so, we propose a shaped comb-allpass (SCAL) filter of the form
\begin{equation}
A(z)=\frac{\alpha\left(1-\beta z^{-1}\right)+z^{-N}}{1-\alpha\left(-\beta z^{-N+1}+z^{-N}\right)}\ ,\label{eq:shaped-comb}
\end{equation}
where $\alpha$ controls the \emph{depth} of the filter and $\beta$
controls the \emph{tilt}. Stability is guaranteed (sufficient condition)
as long as
\begin{equation}
\left|\alpha\right|\left(1+\left|\beta\right|\right)<1\ ,\label{eq:stability-criterion}
\end{equation}
so (\ref{eq:stability-criterion}) can be used to determine the upper
bound on $\alpha$ as a function of $\beta$. The effect of the \emph{tilt}
parameter $\beta$ is demonstrated in Fig. \ref{fig:Effect-of-tilt}
and can be explained by the fact that as $\beta$ increases, the poles
and zeros of the all-pass filter move closer to the unit circle at
high frequencies and away from the unit circle at lower frequencies.

When using a filter of order $N$, there are $N+1$ points on the
frequency axis where the phase response is zero, regardless of $\alpha$.
In other words, there are frequencies where no coherence reduction
occurs. For this reason, we also vary the order $N$ of the filter
so that the ``nulls'' in the phase response change as a function
of time. 

The order $N$ is changed randomly for each new window, subject to
$N_{min}\leq N\leq N_{max}$. We then vary $\alpha$ using
\begin{equation}
\alpha(N)=\min\left(\left(\alpha(N-1)+r_{0}\right),\frac{1-\epsilon}{1+\left|\beta\right|}\right)\ ,\label{eq:random-alpha}
\end{equation}
where $r_{0}$ is a uniformly-distributed random variable chosen in
the $[-r_{max},r_{max}]$ range (typically $r_{max}=0.6$) and $\epsilon\ll1$
controls the distance to the unit circle of the high frequency poles.
The SCAL filter has a delay of $N_{max}=10$ and an overall complexity
of only 23 operations per sample, which is negligible when compared
to the complexity of the adaptive filtering used to cancel the echo.

\section{Psychoacoustically-Masked Noise}

\label{sec:Psychoacoustically-Masked-Noise}

The SCAL processing in Section \ref{sec:Shaped-Comb-Allpass-Filtering}
is mainly effective for frequencies above 2 kHz. For lower frequencies,
the ear is more sensitive to phase distortion (altering stereo image),
so it is preferable to inject noise that is uncorrelated to the audio
signal. In this work, we use the psychoacoustic model from the Vorbis
audio codec, as described in \cite{ValinAES2006}. The output of the
psychoacoustic model determines the amount and spectral shape of the
noise that can be added without significantly altering perceptual
audio quality. The psychoacoustic model is also tuned to introduce
less noise in higher frequencies because those are already decorrelated
by the SCAL filter.

The noise to be added is generated in the frequency domain. Again,
we make use of weighted overlap-and-add to reconstruct the time-domain
signal. To avoid adding a delay to the signal of interest, only the
noise is delayed and is added to the non-delayed input signal. This
can be done without significant audio quality degradation because
of the temporal masking effect. Because we are adding a random signal
during the WOLA process, it is the power that is additive and not
the amplitudes. For that reason, we again need to use a window that
satisfies the Princen-Bradley criterion, even though it is only applied
once. Although a lossy codec could be used \cite{Gansler1998} to
add the noise, it would cause a significant increase in the total
delay (>100 ms with MP3).

\section{Evaluation And Results}

\label{sec:Evaluation-And-Results}

We compare four different coherence reduction algorithms:
\begin{itemize}
\item Proposed algorithm, with shaping and variable order (\textbf{SCAL}),
$\beta=0.43$, $5\leq N\leq10$
\item Proposed algorithm, without shaping or variable order (\textbf{Comb-allpass}),
$\beta=0$, $N=7$
\item Smoothed absolute value (\textbf{smoothed absolute}), $\alpha_{abs}=0.3$
\item First-order, time varying all-pass filter (\textbf{allpass}), $\alpha_{min}=-0.985$
\end{itemize}
The smoothed absolute value non-linearity is included because it was
shown in \cite{Morgan2001} to be among the best memoryless non-linearity.
The time-varying first order all-pass filter is implemented as described
by \cite{Ali1998} but using $\alpha_{min}=-0.985$ to account for
the different sampling rate used in this work.

The block-based phase alteration method proposed in \cite{WU2005}
is excluded from the comparison because the boundary artifacts caused
by the implicit rectangular window causes major quality degradation
at high sampling rate, even for very small amounts of decorrelation.
While a WOLA approach could be used, it would involve additional delay,
something which is not acceptable in this context.

In both the proposed algorithm and the first-order all-pass filter,
there is a random component, so it is possible to independently process
each channel with the algorithm. On the other hand, applying the same
memoryless nonlinearity (smoothed absolute value in this case) to
each channel would not reduce the coherence. For that reason, we invert
the sign of the gain $\alpha_{abs}$ used for each channel.

\subsection{Methodology}

We evaluate the algorithms on eight audio excerpts sampled at 44.1
kHz. Four of the samples are voice samples (male and female speech,
Suzanne Vega, quartet) taken from the EBU Tech 3253 - Sound Quality
Assessment Material (SQAM), while the other four are various music
samples (classical, folk, rock, castanets). 

Because we need the ground truth, all recordings are simulated using
real impulse responses measured in a room with around 220 ms reverberation
time (-60 dB). On the remote end, the samples are played one meter
in front of an XY-stereo microphone (16,384-sample impulse response).
These first recordings are processed using each of the algorithms
in the comparison. The quality of the processed files is evaluated
using the MUltiple Stimuli with Hidden Reference and Anchor (MUSHRA)
\cite{BS1534} methodology\footnote{We used the RateIt graphical interface available at http://rateit.sf.net/}.
Although in the final application the listeners hear the audio through
loudspeakers, we have decided to perform the evaluation using headphones
to make artifacts -- especially stereo image artifacts -- more noticeable.

The processed signals are played on the near end at the back of another
XY-stereo microphone (16,384-sample impulse response) and some background
noise is added to obtain a signal-to-noise ratio (SNR) of 40 dB. Stereo
echo cancellation is then performed using a variant of the MDF algorithm~\cite{Soo1990}
for a 8,192-sample filter length. The filter misalignment is measured
between the filter found by the echo canceller and the first 8192
samples of the real impulse response.

\subsection{Quality and Misalignment}

\begin{figure}
\includegraphics[width=1\columnwidth]{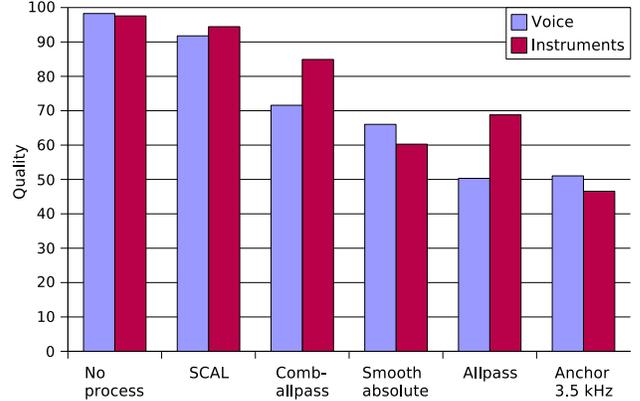}\vspace{-2mm}

\caption{Quality of the signal from different algorithms obtained from a MUSHRA
test (higher is better).\label{fig:Quality-MUSHRA} }
\end{figure}

Fig. \ref{fig:Quality-MUSHRA} shows the results we obtained with
10 listeners\footnote{Results from one additional listener were removed during post-screening
because they were inconsistent (e.g. all algorithms rated 0 including
the reference).} on a MUSHRA quality comparison of the algorithms. We observe that
for both voice and instruments, the quality with the proposed (SCAL)
algorithm is very close to that of the reference (``no process'')
and significantly higher than all other algorithms, with greater than
99\% confidence on a permutation test. This also shows that the shaped
aspect (parameter $\beta$) of the comb-allpass filter is important,
since the simple comb-allpass filter has significantly lower quality.

\begin{figure}
\includegraphics[width=1\columnwidth]{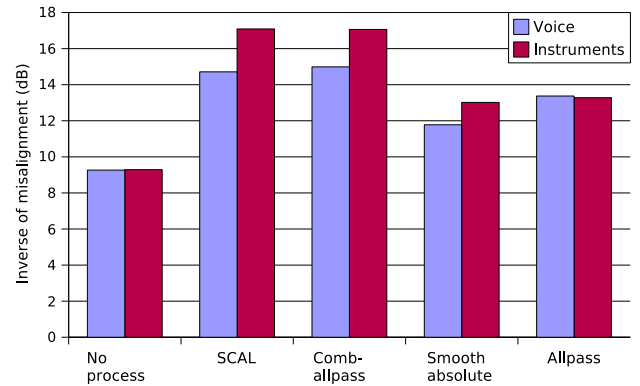}\vspace{-2mm}

\caption{Inverse of the filter misalignment (higher is better).\label{fig:Inverse-misalignment}}
\vspace{-4mm}
\end{figure}

Fig. \ref{fig:Inverse-misalignment} shows the misalignment obtained
using each algorithm after 10 seconds. We can observe that convergence
with the proposed method is a significant improvement over the smoothed
absolute value, the first-order allpass filter and the unprocessed
signal. The total misalignment is approximately the same as with the
simple comb-allpass filter. 

\begin{figure}
\includegraphics[width=1\columnwidth]{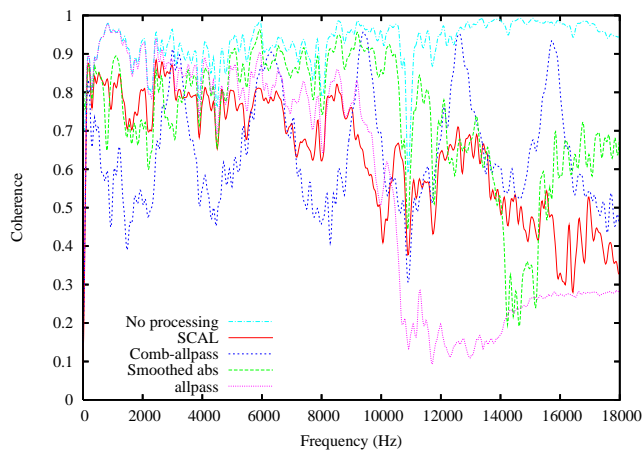}\vspace{-2mm}

\caption{Inter-channel coherence after processing (lower is better)\label{fig:Coherence}}
\vspace{-4mm}
\end{figure}

Lastly, the average inter-channel coherence is shown in Fig. \ref{fig:Coherence}
as a function of frequency for a male speech sample. We see that the
proposed algorithm is more constant than other algorithms. It can
also be observed that for the comb-allpass filter with a constant
order some frequencies are still highly correlated. This is due to
``nulls'' in the phase response that do not depend on the value
of $\alpha$.

\subsection{Qualifying Algorithm Artifacts}

Listening to the samples makes it possible to qualify the artifacts
caused by the algorithms under test and make the following remarks.
\vspace{-2mm}

\subsubsection*{Proposed algorithm}

Some listeners reported a mild ``flanging'' effect when listening
to the proposed algorithm, along with slight movement of the stereo
image from left to right. Overall, the artifacts are less severe than
those of the other algorithms and the stereo image distortion would
likely become less noticeable when listening with loudspeakers. \vspace{-2mm}

\subsubsection*{Proposed algorithm, without shaping and variable order}

When we remove the shaping of the comb-allpass filter, the flanging
effect becomes worse and the distortion in stereo image becomes much
more noticeable. This is due to increased changes in phase below 1
kHz, which affects the interaural phase difference (IPD). That degradation
is most noticeable on the well known Suzanne Vega \emph{a cappella}
excerpt.

\vspace{-2mm}

\subsubsection*{Smoothed absolute value}

Being a non-linear function, the smoothed absolute value produces
inter-modulation distortion. On harmonic signals such as speech, the
distortion is mostly masked and is perceived as additional ``harshness''.
Another artifact can be observed in the castanets sample. Because
castanets produce strongly asymmetric time-domain impulses, the smoothed
absolute value causes one of the channels to be amplified more than
the other, resulting in a very disturbing ``bouncing'' stereo image.
On tonal non-harmonic signals such as the glockenspiel (not included
in the formal evaluation), the inter-modulation distortion effect
can cause new tones to appear in some regions of the spectrum. Some
tones even appear at low frequencies, which has the effect of changing
the perceived fundamental frequency. \vspace{-2mm}

\subsubsection*{First-order all-pass filter}

The main artifact introduced by the first-order all-pass filter is
a nearly white crackling noise that is the result of varying the filter
coefficient $\alpha$ from one sample to another. For most samples,
the noise is masked at lower frequency, so it is usually perceived
as a high-frequency crackling noise. It is mainly perceivable on very
tonal samples, that do not leave much room for masking noise components.

\section{Conclusion}

\label{sec:Conclusion}

In this paper, we have demonstrated that it is possible to reduce
the coherence between the left and right channels in a video-conference
application without significantly reducing the audio quality. The
proposed method includes a shaped comb-allpass (SCAL) filter to reduce
coherence at higher frequencies and psychoacoustically masked noise
injection at lower frequencies. Novel aspects of this work include
the shaping of the phase alteration to better match human stereo perception,
as well as the use of windowing the allpass filter output to prevent
blocking artifacts.

The proposed method was shown to outperform other existing methods
both in terms of quality and amount of decorrelation provided, leading
to better echo cancellation results. Moreover, the total complexity
of the proposed algorithm is kept small so that it does not significantly
increase the complexity of a complete echo cancellation system.

\bibliographystyle{IEEEbib}
\bibliography{echo}

\end{document}